\documentstyle[12pt]{article}
\def\fun#1#2{\lower3.6pt\vbox{\baselineskip0pt\lineskip.9pt
\ialign{$\mathsurround=0pt#1\hfil##\hfil$\crcr#2\crcr\sim\crcr}}}

\newcommand{\be}{\begin{equation}}
\newcommand{\ee}{\end{equation}}
\newcommand{\bd}{\begin{displaymath}}
\newcommand{\ed}{\end{displaymath}}
\newcommand{\ba}{\begin{array}}
\newcommand{\ea}{\end{array}}
\newcommand{\bt}{\begin{tabular}}
\newcommand{\et}{\end{tabular}}
\newcommand{\bc}{\begin{center}}
\newcommand{\ec}{\end{center}}

\begin{document}


\hfill {\large\bf E2--97--325}

\hfill {\large\bf November 1997}

\begin{center}
{\large\bf Charmed sea contribution to the inclusive 
hadroproduction \\ of the mesons with open charm 
in the Quark -- Gluon String Model}\\[0.5cm]
G.H.Arakelyan \footnote{Yerevan Physics Institute,
 Yerevan 375036, Armenia \\E--mail:argev@vxitep.itep.ru~~~~
argev@nusun.jinr.ru} \\[0.5cm]
Joint Institute for Nuclear Research, Dubna, Russia \\[0.5cm]
\end{center}


\vspace{1.5cm}

\begin{abstract}

The hadroproduction of charmed mesons is discussed within the
framework of the modified Quark-Gluon String model (QGSM) taking
into account the decays of corresponding $S$--wave resonances
like $D^*$. It is shown that the estimate of the $c\bar{c}$--pair
contribution to the quark sea of colliding hadrons strongly
depends on the parametrization of the function of the leading
light quark fragmentation into charmed mesons. A description of
the existing experimental data on inclusive spectra and
asymmetries of leading and nonleading $D$ - and $D^*$ - meson
production in $ \pi p$ -, $p p$ - and $\Sigma p$ - collisions is
obtained.  The predictions for leading/nonleading asymmetry in
$\Sigma^- p$-- and $\Xi^-p$-- collisions are also given.

\end{abstract}

\vfill

\noindent

\newpage

\section{Introduction}

~~~~~Hadroproduction of charmed particles is now being investigated
in many experiments at various energies of the different initial
hadron beams. One of the most striking features of charm hadroproduction
is the leading particle effect \cite{AP}.
For example, in $ \pi^- (d \bar u)$ interactions with hadrons and nuclei
more $D^-(d \bar c)$ than $D^+(c \bar d)$ are produced at large $x_F$.

Leading charm production can be quantified by studies of the
production asymmetries between leading and nonleading charm
production. The good measurements of the differential cross
sections up to high value of Feynman $x_F \rightarrow 1$ gives a
possibility of obtaining the lea\-ding/non\-lea\-ding asymmetry
defined as

\be
\label{1}
A({D^-,D^+})=\frac{\sigma(D^-) - \sigma(D^+)}
{\sigma(D^-) + \sigma(D^+)}
\ee

The experimental data WA82 \cite{ADPL}, E769 \cite{ALPRL72},
E791 \cite{AIPL371} and WA92 \cite{WA92PPE} show that the asymmetry
increases
from nearly zero at small $x_F$ to 0.5 around $x_F=0.6$ and does
not show strong energy dependence in the range
$p_L=250 - 500GeV/c$. This means that leading charm asymmetry is
primerily located at large $x_F$. Thus the asymmetry $A$ reflects the
physics at only a small fraction of the total $D^{ \pm}$ cross section.

Neutral D's were usually not used in the analysis since they can also be
produced by the decay of nonleading $D^{* \pm}$ -- mesons. Recently
the WA92 (Beatrice) collaboration \cite{WA92PPE} presented the data on
$A{(D^0,\bar D^0)}$ asymmetry at $350\;GeV/c$.

How can one explain the origin of leading charm asymmetry within the
context of different theoretical approaches?

Perturbative QCD at leading order predicts that c and $\bar c$ quarks are
produced with the same distributions. The asymmetry in this case is
equal to zero.

In PYTHIA, a Monte Carlo program based on the Lund string fragmentation
model \cite{BS87,PY57}, it is assumed that heavy quarks
are produced in the initial state with a relatively small longitudal
momentum fraction by the leading twist fusion processes. In order to
produce a strong leading particle effect at large $x_F$, the string has
to accelerate the heavy quark as it fragments and to form the final heavy
hadron.

In papers \cite{BRPL93,VBH,VB1,VB2} a QCD mechanism which produced a
strong lea\-ding/non\-lea\-ding asymmetry at large $x_F$ is discussed.
The basic assumption of the intrinsic charm model \cite{VBH,VB1,VB2}
is the coalescence of charmed quarks and a projectile valence quarks
occur in the initial state. For example, $ \pi^-$ can fluctuate into a
$ \mid \bar u dc \bar c>$ Fock state. These states are produced in
QCD from amplitudes involving two or more gluons, attached to the
charmed quarks. Since the charm and valence quarks have the same rapidity
in these states, it is easy for them to coalesce into charmed hadrons
and produce leading particle correlation at large $x_F$ where this
mechanism can dominate the production rate.

The leading effect was also considered in different recombination
models \cite{ZMU,HWA,LIK,BED}, where in addition to processes of
$c$--quark hadronization, the recombination of the light valence
quark with the $c$--quark is taken into account.

In the framework of the Quark--Gluon String model (QGSM)
\cite{KTM82,AK82,KTM84} the "intrinsic" $c \bar c$ pairs can be
taken into account only as a small admixture in the quark sea of
colliding particles. This assumption is in some disagreement
with the IC hypothesis \cite{VBH,VB1,VB2} because $c$ and $\bar c$
quarks should rather be considered as valence quarks.

The first attempt to introduce the charmed sea contribution into
the QGSM was made in \cite{ASH} without comparison with the
experimental data on leading/nonleading asymmetry.

The qualitative analysis of the charmed sea contribution in
the framework of the QGSM was made in \cite{PY60}, where the term
 "topological charm" for this phenomenon was proposed in order to
distinguish it from "intrinsic charm" suggested by Brodsky and Vogt
\cite{VBH,VB1,VB2}.

The QGSM prediction for the inclusive spectra of
charmed hadrons was considered (without intrinsic charm) in
\cite{KPY43} - \cite{AVC}.

In the present paper we analysed the experimental data on
leading/nonleading $D$-meson production asymmetry and inclusive
cross section in the framefork of the modification of the QGSM,
developed in \cite{AVZA,AVC} taking into account the charmed
sea contribution.

\section{Model description}

~~~~~ The modification of the QGSM taking into account the
contributions from decays of corresponding $S$--wave resonances
was developed in \cite{AVZA}. Under this assumption the invariant
cross section of the produced hadron $h$ integrated over transverse
momenta $p_{\bot}$ can be written as

\be
\label{2}
x\frac{d\sigma^h}{dx}=x\frac{d\sigma^{h{dir}}}{dx}+ \sum_{R}
\int\limits^{x^*_+}_{x^*_-}x_R\frac{d\sigma^R}{dx_R}\Phi(x_R)dx_R~.
\ee

Here, $x\frac{d\sigma^{h{dir}}}{dx}$ is the direct production
cross section of the hadron $h$ and $x_R\frac{d\sigma^R}{dx_R}$ is
the $R$--resonance production cross section. The function $\Phi(x_R)$
describes the decay of the resonance $R$ into hadron $h$. $S$--wave
charmed resonances decay into stable charmed
particles emitting a $\pi$--meson or a $\gamma$--quantum \cite{RPP},
and we describe the two--body kinematics of this decay as in \cite{ASHK}.
After integration over the transverse momenta of both the hadron $h$ and
the resonance $R$ the function $\Phi(x_R)$ has the form

\be
\label{3}
\Phi(x_R)=\frac{M_R}{2p^*}\frac{1}{x^2_R}~.
\ee
In eqs. (\ref{2}) and (\ref{3}) $x_R$ is the Feynman variable
of the resonance $R$,

\bd
x^*_+=\frac{M_R\tilde x}{E^*-p^*},\quad x^*_-=\frac{M_R\tilde
x}{E^*+p^*},\quad \tilde x=\sqrt{x^2+x^2_{\bot}},\quad
x_{\bot}=\frac{2\sqrt{<p^2_{\bot}>+m^2}}{\sqrt s},
\ed

$m$ is the mass of the produced hadron $h$, $M_R$ is the mass of the
resonance, $E^*$ and $p^*$ are energy and $3$--momentum of hadron $h$
in the resonance rest frame, $<p^2_{\bot}>$ is the average transverse
momentum squared of the hadron $h$.

The inclusive spectra of the hadron $h$ in the framework of the QGSM
has the form

\be
\label{4}
x\frac{d\sigma^h}{dx}=\sum^{\infty}_{n=0}\sigma_n(s)\varphi^h_n(x)~,
\ee

where $\sigma_n(x)$ is the cross section of $n$--pomeron shower
production and $\varphi^h_n(x)$ determines the contribution of
the diagram with $n$ cut pomerons. The expressions for
$\sigma_n(s)$ and the corresponding parameter values
for $pp$ and $\pi p$ collisions are given in  \cite{KPY43}--\cite{AVZA}.

The function $\varphi^h_n(x)$ $(n>1)$ for $\pi p$ interaction can be
written in the form  \cite{SHY44}--\cite{AVZA}

\bc

\begin{eqnarray}
\label{5}
\varphi^h_n(x)=a_0^D(f^h_{\bar q}(x_+,n)f^h_q(x_-,n)+
f^h_q(x_+,n)f^h_{qq}(x_-,n) \nonumber  \\
+2(n-1)f^h_{sea}(x_+,n)f^h_{sea}(x_-,n))
\end{eqnarray}

\ec

and for baryon--proton interaction

\bc

\begin{eqnarray}
\label{6}
\varphi^h_n(x)=a_0^D(f^h_{qq}(x_+,n)f^h_q(x_-,n)+f^h_q(x_+,n)
f^h_{qq}(x_-,n)+ \nonumber  \\
2(n-1)f^h_{sea}(x_+,n)f^h_{sea}(x_-,n))~,
\end{eqnarray}

\ec

where $x_{\pm}=\frac{1}{2}[\sqrt{x^2+x^2_{\bot}}\pm x]$.

The functions $f^h_i(x,n)(i=q,\bar q,qq,q_{sea})$ in (\ref{5}) and
(\ref{6}) describe the contributions of the valence/sea quarks,
antiquarks and diquarks, respectively and were determined by
convolution of the corresponding distribution functions $u_i(x,n)$ in
the colliding hadrons with the function of quark/diquark
fragmentation into hadron $h$ $G^h_i(x,n)$ \cite{AVZA}:

\be
\label{8}
f_i(x,n)=\int^1_x u_i(x_1,n)G_i(x/x_1)dx_1~.
\ee

The projectile (target) contribution depends on the variable $x_+$
$(x_-)$.

The functions $f^h_{sea}(x_+,n)$ and $f^h_{sea}(x_-,n)$ in (\ref{5})
and (\ref{6}) are parametrized in the form

\newpage

\begin{eqnarray}
\label{7}
f_{sea}^{h} (x_{\pm},n) & = & \frac{1}{2 + \delta_{s} + \delta_{c}}
\left[ \int_{x_{\pm}}^{1}
u_{\overline{u}} (x_{1},n) \frac{G_{\overline{u}}^{h} (x_{\pm} / x_{1}) +
G_{u}^{h} (x_{\pm} / x_{1})}{2} dx_{1} \right. \nonumber  \\
& + &  \left. \int_{x_{\pm}}^{1} u_{\overline{d}} (x_{1},n)
\frac{G_{\overline{d}}^{h} (x_{\pm} / x_{1}) +
G_{d}^{h} (x_{\pm} / x_{1})}{2} dx_{1} \right. \nonumber  \\
& + &  \left. \delta_{s} \int_{x_{\pm}}^{1} u_{\overline{s}} (x_{1},n)
\frac{G_{\overline{s}}^{h} (x_{\pm} / x_{1}) + G_{s}^{h} (x_{\pm} / x_{1})}
{2} dx_{1} \right. \nonumber \\
& + &  \left. \delta_{c} \int_{x_{\pm}}^{1} u_{\overline{c}} (x_{1},n)
\frac{G_{\overline{c}}^{h} (x_{\pm} / x_{1}) + G_{c}^{h} (x_{\pm} / x_{1})}
{2} dx_{1} \right] \;.
\end{eqnarray}

The parameters $\delta_{s} \sim 0.2-0.3$ and $\delta_{á}$ stand for
strange and charmed quark supression in the sea. The value of
$\delta_{c}$ will be found later from the comparison with the
experimental data.

A full list of the quark/diquark distribution functions in the
$\pi$--meson, $p$, $\Sigma^-$, and $\Xi^-$--hyperons, used in this
work, is given in \cite{AVZA}.

We assume the following parametrization of the charmed sea in different
hadrons

\be
\label{9}
u^{\pi}_c(x,n)=C^{\pi}_cx^{-\alpha_{\psi}(0)}(1-x)^{\gamma^{\pi}_c
+2(n-1)(1-\alpha^0_{\rho})}~,
\ee

$$\gamma^{\pi}_c=-\alpha_{\rho}(0)+
(\alpha_{\rho}(0)-\alpha_{\psi}(0)).$$

\be
\label{10}
u^{p}_c(x,n)=C^p_cx^{-\alpha_{\psi}(0)}(1-x)^{\gamma^p_c+
2(n-1)(1-\alpha_{\rho}(0))}~,
\ee

$$\gamma^p_c=(\alpha_{\rho}(0)-2\alpha_{N}(0))+
(\alpha_{\rho}(0)-\alpha_{\psi}(0)).$$

\be
\label{11}
u^{\Sigma^-}_c(x,n)=C^{\Sigma^-}_cx^{-\alpha_{\psi}(0)}
(1-x)^{\gamma^{\Sigma^-}_c+2(n-1)(1-\alpha_{\rho}(0))}~,
\ee

$$\gamma^{\Sigma^-}_c=(\alpha_{\rho}(0)-2\alpha_{N}(0))+
(\alpha_{\rho}(0)-\alpha_{\phi}(0))+
(\alpha_{\rho}(0)-\alpha_{\psi}(0)).$$

\be
\label{12}
u^{\Xi^-}_c(x,n)=C^{\Xi^-}_cx^{-\alpha_{\psi}(0)}
(1-x)^{\gamma^{\Xi^-}_c+2(n-1)(1-\alpha_{\rho}(0))}~,
\ee

$$\gamma^{\Xi^-}_c=(\alpha_{\rho}(0)-2\alpha_{N}(0))+
(\alpha_{\rho}(0)-\alpha_{\phi}(0))+
2(\alpha_{\rho}(0)-\alpha_{\psi}(0)).$$

 The coefficients
$C_c$ in (\ref{9} -- \ref{12}) are determined by the normalization
condition $$\int^1_0u^n_i(x,n)dx=1$$.

Further we will assume that the fragmentation functions of quarks
and diquarks do not depend on the spin of the picked up quark (or
diquark) \cite{AVZA}. This leads to the equality of the fragmentation
functions of the corresponding quarks or diquarks into different $D$
and $D^*$ mesons. In the present paper we use the parametrization of
the light quark and diquark fragmentation function given in
\cite{KPY43,PY56} but with other values of free parameters.

The nonleading fragmentation functions of the quark into $D$-- mesons
have the form

\be
\label{13}
G_{d}^{D^{+}}(z)=(1-z)^{-\alpha_{\psi}(0)+\lambda+
2(1-\alpha_{R}(0))},
\ee

where  $\lambda$=2$\alpha_{D^{*}}^{\prime}$(0)$\bar{p_{\bot D^{*}}^{2}}$,
$\alpha_{\psi}(0)=-2.18$ \cite{AVZA}.

The leading type fragmentation function is parametrized as:

\be
\label{14}
{G}_{d}^{D^{-}}(z)=(1-z)^{-\alpha_{\psi}(0)+\lambda}
(1+a_{1}^{D}z^{2}).
\ee

The $a_{1}z^{2}$ term was introduced in \cite{KPY43} by analogy with
the fragmentation function into the $K$--meson \cite{KFF} and stands
for the density of mesons at the end of the fragmenting quark--gluon
string. The value of $a_1$ is poorly determined from the
inclusive cross section data (see \cite{PY60} -- \cite{SHY44}). In
our approach it is possible to have reasonable agreement with the
asymmetry data without introducing the charmed sea contribution
with $a_1=2$. A bit better agreement can be reached by introducing
a small fraction of charmed sea( see the next section).
Introduction of the term ~$1+z^2$ in (\ref{14}) is rather arbitrary.
We also consider another parametrization

\be
\label{15}
{G}_{d}^{D^{-}}(z)=(1-z)^{-\alpha_{\psi}(0)+\lambda}
(1+a_{1}^{D}z^{4}).
\ee

The fragmentation function of the $c$($ \bar c$) -- quark into
charmed mesons was parametrized in the form

\begin{equation}
\label{16}
G_{c(\bar{c})}^{D^{-}(D^{+})}(z)=\frac{b^D}{a^D_0}z^{1-\alpha_{\psi}(0)}
(1-z)^{\lambda-\alpha_R(0)}.
\end{equation}
where $b^D=1$.

The value of $\delta_c$ is different for different parametrizations
and was obtained from the comparison with the experimental data
\cite{ADPL} - \cite{WA92PPE},\cite{ABPL161} - \cite{ABPL169}.

For the hyperon beam processes it is needed to know the fragmentation
functions of the strange quark and the corresponding diquarks. According
to \cite{KFF} the parametrization was chosen as

\be
\label{17}
{G}_{s}^{D}(z)=(1-z)^{\lambda-\alpha_{\psi}(0)-\alpha_R(0)
-\alpha_{\phi}(0)+2}
\ee

\be
\label{18}
{G}_{ds}^{D^-}(z)=(1-z)^{\lambda+1-\alpha_{\psi}(0)+\alpha_R(0)
-2\alpha_N(0)}(\frac{1+a_1z^2}{2}+\frac{(1-z)^2}{2})
\ee

\be
\label{19}
{G}_{ds}^{D^+}(z)=(1-z)^{\lambda-\alpha_{\psi}(0)+2(1
-\alpha_N(0)}(\frac{1}{2}+\frac{(1-z)^2}{2})
\ee

\be
\label{20}
{G}_{ss}^{D}(z)=(1-z)^{\lambda-\alpha_{\psi}(0)+2(1-\alpha_N(0))}
\ee
The values of the parameters will be given in the next section.

\section{Comparison with the experimental data and the predictions
of the model}

~~~~~In this section, we consider the description of the existing
experimental data for the  $D$ and $D^*$--meson hadroproproduction
in the framework of the present model using fragmentation functions
(\ref{14}) and (\ref{15}).
In what follows, we will present the three curves in all figures.
They correspond to three versions under consideration: 1) $d$--quark
fragmentation function into $D^-$--meson (\ref{14}) with the
parameters $a_0^D=0.0007$, $a_1^D=2$,
without taking into account intrinsic charm ($\delta_c=0.$).
This curve is shown by full line and marked by number 1;
2) the same as in 1) but $\delta_c=0.005$, dashed line, marked by 2;
3) $d$--quark fragmentation function into the $D^-$--meson according
(\ref{15}) with the parameters $a_0^D=0.0007$, $a_1^D=8$,
$\delta_c=0.01$, dash-dotted line, number 3. All sets of parameters
were determined from the comparision with the experimental data
\cite{ADPL} - \cite{WA92PPE} and \cite{ABPL161} - \cite{ALPRD49}.
All theoretical curves in the model under consideration are
sums of the directly produced $D$--meson cross section and the
contribution of the decay of the corresponding $D^*$ resonanse.

The experimental data on the $x_{F}$ dependence of  €($D^-$,$D^+$)
asymmetry measured by different groups, WA82 for
$P_L=250\;GeV/c$ \cite{ADPL}, E769 for $P_L=340\;GeV/c$
\cite{ALPRL72}, and the recent data of WA92 (Beatrice
collaboration) for $P_L=350\;GeV/c$ \cite{WA92PPE} are presented
in Fig.1. The theoretical curves were calculated for $P_L=340\;Ē'/c$.

The Beatrice collaboration presented the measurements of
€($D^0$,$\bar D^0$) asymmetry \cite{WA92PPE} . This data
are compared with our calculations in Fig.2. In Fig.3 we plotted
the experimental points for the asymmetry €($D^-$,$D^+$) at
$P_L=500\;GeV/c$ \cite{AIPL371} together with the QGSM calculations
in the same kinematic region.

The experimental data on the $x_F$-- dependence of the inclusive
distributions of all $D$--mesons in  $\pi^-p$ and $pp$ interactions
at $P_L=200\;GeV$ \cite{BZC39} and $250\;GeV/c$ \cite{ALPRL77}
 are presented in Figs.4 and 5 respectively.
The theoretical curves are calculated for the sums of the spectra
of all $D$--mesons at $P_L=250GeV/c$.

The model calculations for the spectra of the sum of all $D$-- mesons
in the reaction $\pi^- p \rightarrow D X$ at $P_L=360\;GeV/c$
are compared with the experimental data at $P_L=350\;GeV/c$
\cite{WA92PPE} and $P_L=360\;GeV/c$ \cite{ABPL161,ABZC31} in Fig.6.

As far as we  consider here the QGSM spectra of resonances taking
into account their subsequent decays, it is important to compare our
calculations with the available data on $D^*$-- meson production.
The data on the reactions $\pi^-p\to D^{*+}/D^{*-}X$ and
$\pi^-p\to D^{*0}/\bar D^{*0}X$ at $360\;GeV/c$ \cite{ABPL169} are
compared with our predictions in Fig.7 and 8 respectively.

In \cite{ALPRD49} they presented the asymmetry of the leading
 ($D^{*-},\bar D^{*0}$) and nonleading ($D^{*+},D^{*0}$) vector
resonances with open charm in  $\pi^- p$-- collisions, integrated
in the $x_F>0$ region. This value is $A(L,NL)=0.09 \pm 0.06$.
The calculations of the model give the following values for all
parametrizations used in this work:
1 -- $A(L,NL)=0.103$, 2 -- $A(L,NL)=0.097$, 3 -- $A(L,NL)=0.088$.
As we can see the model calculation agrees with the experimental
measurements.  The data on the $x_F$ dependence of differential
cross section of the reactions $\pi^-p\to D^{*+}/D^{*-}X$ and
$\pi^-p\to D^{*0}/\bar D^{*0}X$ at $360\;GeV/c$\cite{ABPL169}
are compared with our calculations in Fig.7 and 8.
These results confirm our assumption about equality of the spectra of
the vector ($D^*$) and directly produced pseudoscalar ($D$) mesons.

The predictions of the model for the inclusive
spectra of  $D^{-}$ and $D^{+}$ mesons in $\Sigma^-p$ collision
at $330\;GeV/c$ (WA89) are given in Fig.9 and 10.

The predictions for the $x_F$-- dependence of the asymmetry
€($D^-$,$D^+$) for $\Sigma^- p$ and $\Xi^- p$-- collisions are
presented in Figs. 10 and 11, respectively.

As we can see, the existing data do not allow to have definite
conclusion on the charmed sea contribution in the QGSM. The curves
in Figs. 1-3 show that it is possible to have a reasonable
description of the data on leading/nonleading asymmetry without
charmed sea contribution if one chooses the value of the
corresponding parameter $a_1=2$ (full line in the figures). The
charmed sea contribution is noticeable in the region $x_F
\rightarrow 1$, and negligible for low $x_F$. This is due to
large suppression of the charmed sea at $x_F \rightarrow 0$
($x_F^{1-\alpha_{\psi}(0)}$).

In \cite{PY60} the asymmetry behaviour
for another set of parameters ($a_1 =4$, $\delta_c =0.01a_0$)
was compared with the calculation for the old one \cite{KPY43,PY56},
which gives larger values of the asymmetry for the entire $x_F$
interval.

The comparison of the experimental data with the calculations using
parametrizations of fragmentation function (\ref{15})
shows that in this case agreement is even slightly better.

The main conclusion from the results considered is that the
existing data do not allow an unambiguos quantitative
answer conserning the charmed sea contribution in the QGSM.

{\it Acknowledgments}. The author is grateful to~ A.B.Kaidalov,~
K.G.Boreskov,~ \\ E.Chu\-da\-kov and O.I.Piskounova for useful
discussion. This work was supported by Grant INTAS--93-0079ext
and Armenian Sciense Foundation (Grant 94-681).

\newpage

\section*{Figure captions}

\noindent
\bt{p{1.5cm}p{14.5cm}}

Fig.1 & Comparison of the QGSM calculations with the experimental
data on $D^{-}/D^{+}$ at
$P_L=250\;GeV/c$ \cite{ADPL} and $P_L=340\;GeV/c$ \cite{ALPRL72} and
$P_L=350\;GeV/c$ \cite{WA92PPE}. The theoretical curves were calculated
for $P_L=340\;GeV/c$, full curve (1) without including charmed sea;
dashed (2) with taking into account the charmed sea; dash-dotted curve
(3) $d$-- quark fragmentation function according the formulae (\ref{15}).
\\[2mm]

Fig.2 & Comparison of the QGSM calculations with the experimental data on
$D^{0}/{\bar D^{0}}$ asymmetry at $P_L=350\;GeV/c$ \cite{WA92PPE}.
The curves are the same as in fig.1. \\[2mm]

Fig.3 & Comparison of the QGSM calculations with the experimental data on
$D^{-}/D^{+}$ asymmetry at $P_L=500\;GeV/c$ \cite{AIPL371}.
The curves are the same as in fig.1. \\[2mm]

Fig.4 & Differential cross sections of the $D$--meson production
in $\pi^-p$-- interaction. The experimental data are  $200\;GeV/á$
(NA32) \cite{BZC39}, $250\;GeV/c$ (E769) \cite{ALPRL77}.
The curves are the same as in fig.1. \\[2mm]

Fig.5 & Differential cross sections of the $D$--meson production
in $pp$-- interaction. The experimental data are  $200\;GeV/á$
(NA32) \cite{BZC39}, $250\;GeV/c$ (E769) \cite{ALPRL77}.
The curves are the same as in fig.1. \\[2mm]

Fig.6 & Differential cross sections of the $D$--meson production
in $\pi^-p$-- interaction. The experimental data are  $360\;GeV/á$ (NA27)
\cite{ABPL161,ABZC31}, $350\;GeV/c$ (WA92) \cite{WA92PPE}.
The curves are the same as in fig.1. \\[2mm]

Fig.7 & The $x_F$--dependence of $D^{*+}/D^{*-}$--mesons
in $\pi^-p$ interaction at $360\;GeV/c$ \cite{ABPL169}.
The curves are the same as in fig.1. \\[2mm]
Fig.8 & The $x_F$--dependence of $D^{*0}/\bar D^{*0}$--mesons
in $\pi^-p$ interaction at $360\;GeV/c$ \cite{ABPL169}.
The curves are the same as in fig.1. \\[2mm]

Fig.9 &Model prediction for the $x_F$--dependence of the
$D^{-}$--meson cross section in $\Sigma^-p$ interaction at
$330\;GeV/c$. The curves are the same as in fig.1. \\[2mm]

Fig.10 &Model prediction for the $x_F$--dependence of the
$D^{+}$--meson cross section in $\Sigma^-p$ interaction at
$330\;GeV/c$. The curves are the same as in fig.1. \\[2mm]

Fig.11 & Predictions for the $x_F$--dependence of $D^-/D^+$
asymmetry for the  $\Sigma^-$ beam at $330\;GeV/c$.
The curves are the same as in fig.1. \\[2mm]

Fig.12 & Predictions for $x_F$--dependence for the $D^-/D^+$ asymmetry
on $\Xi^-$ beam at $330\;GeV/c$.
The curves are the same as in fig.1. \\[2mm]

\et

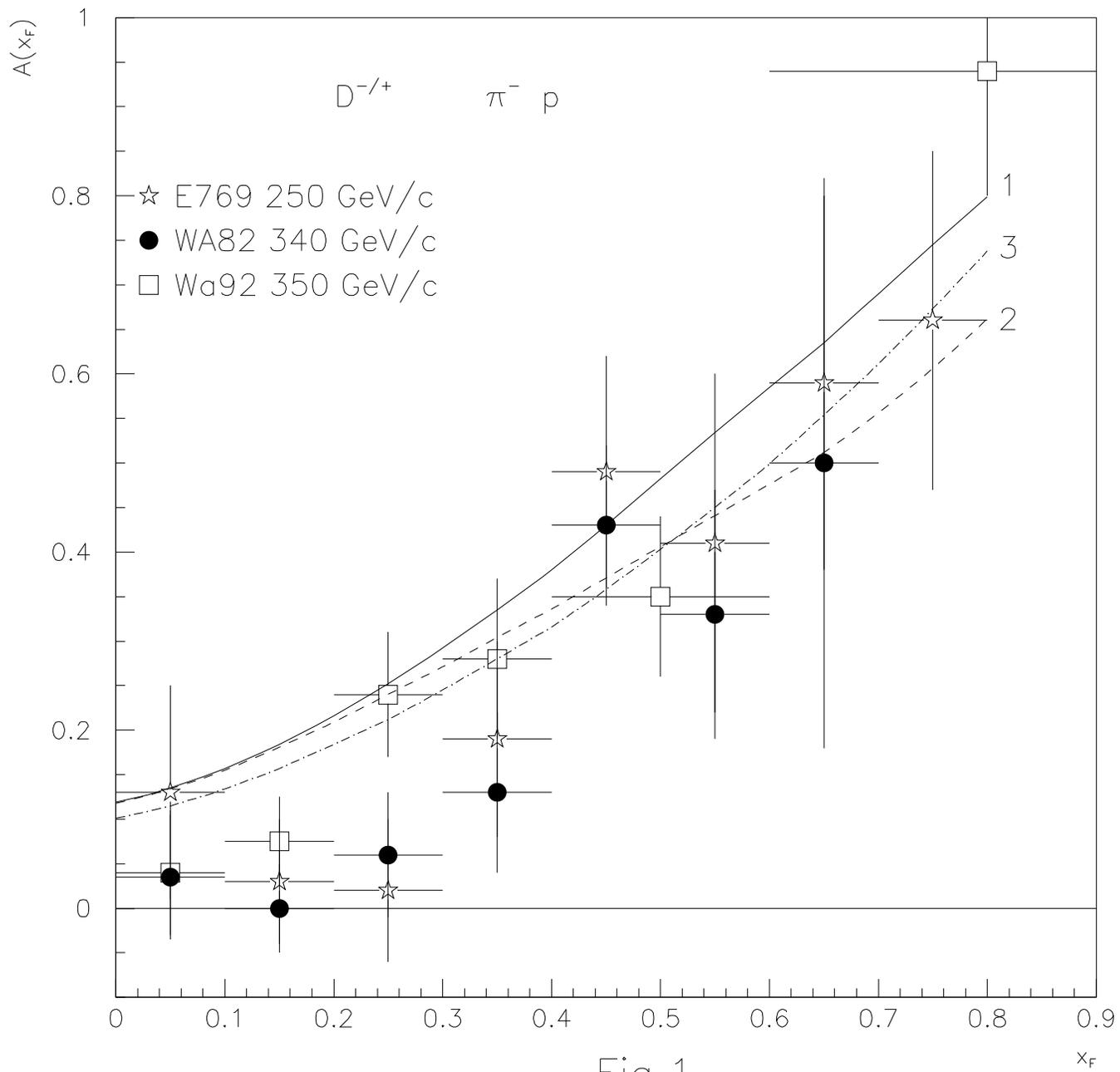
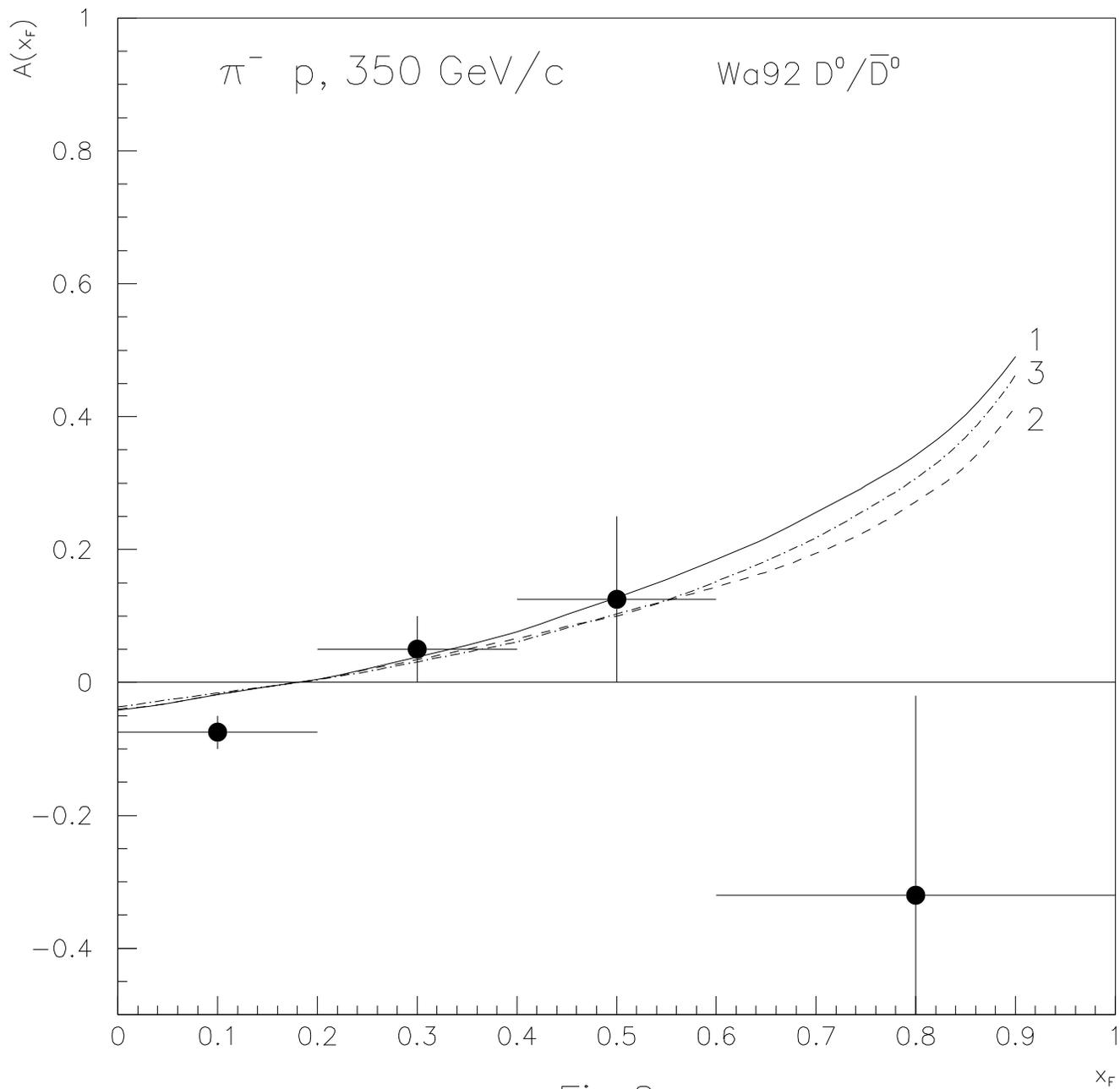
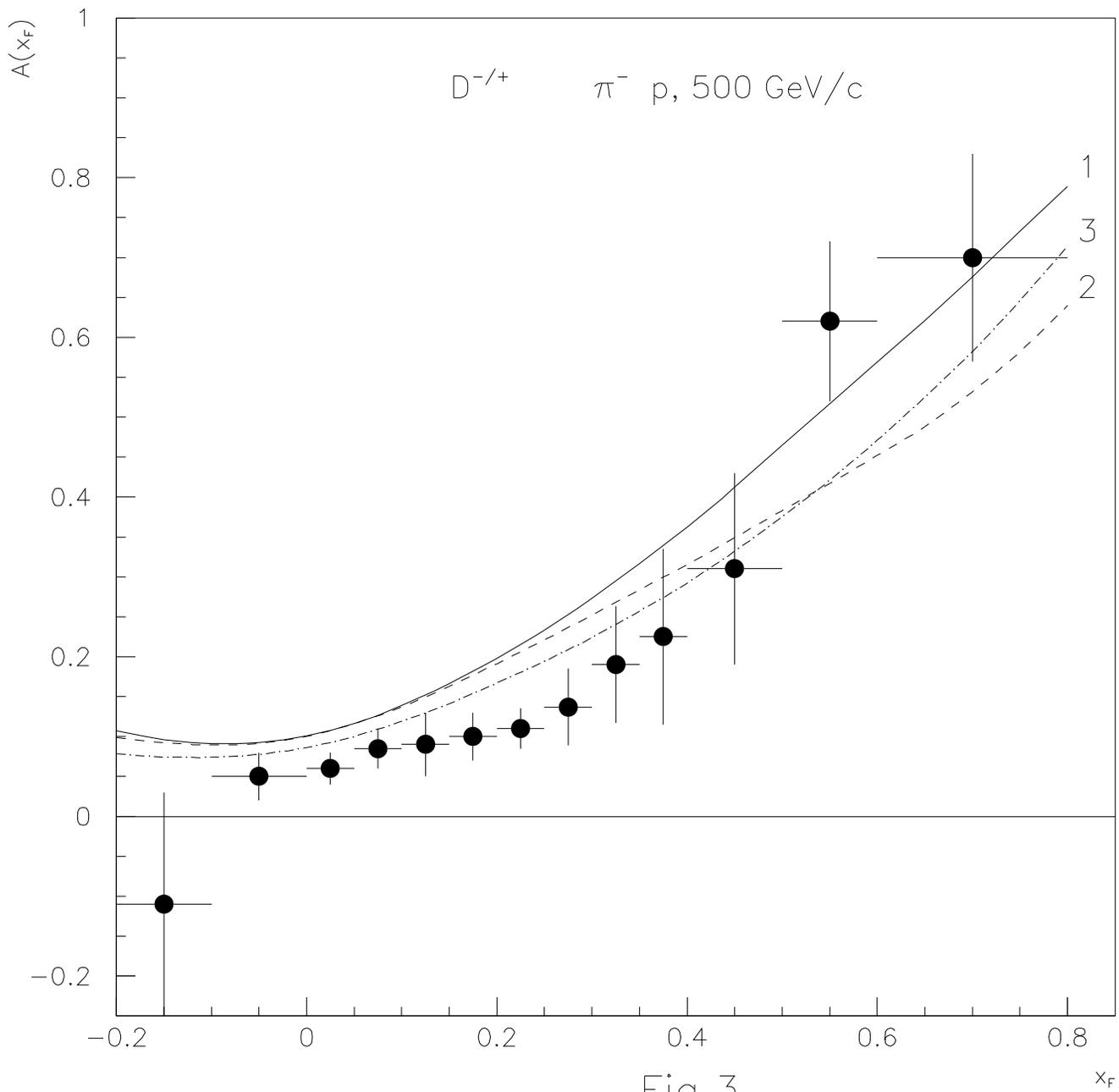
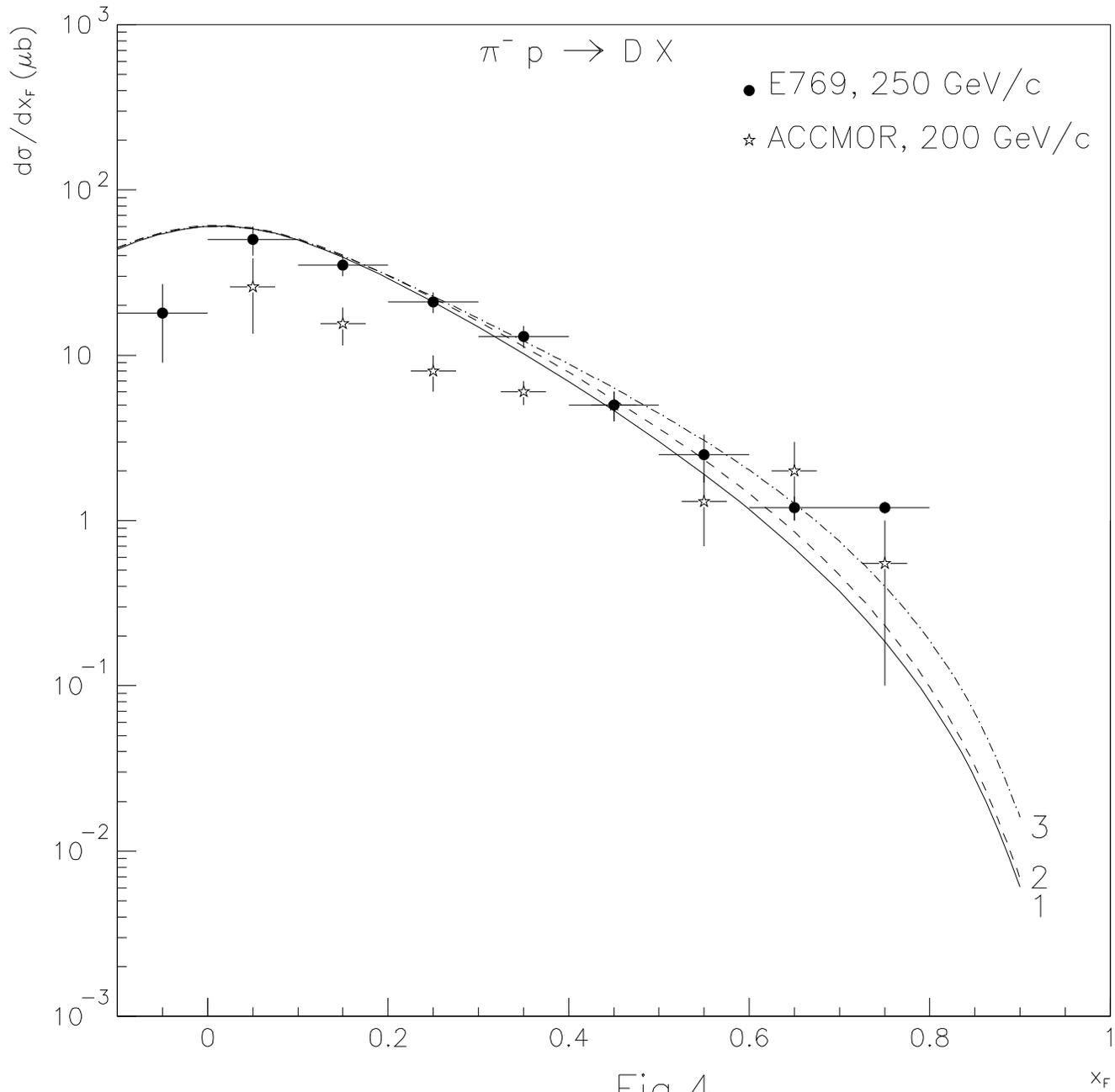
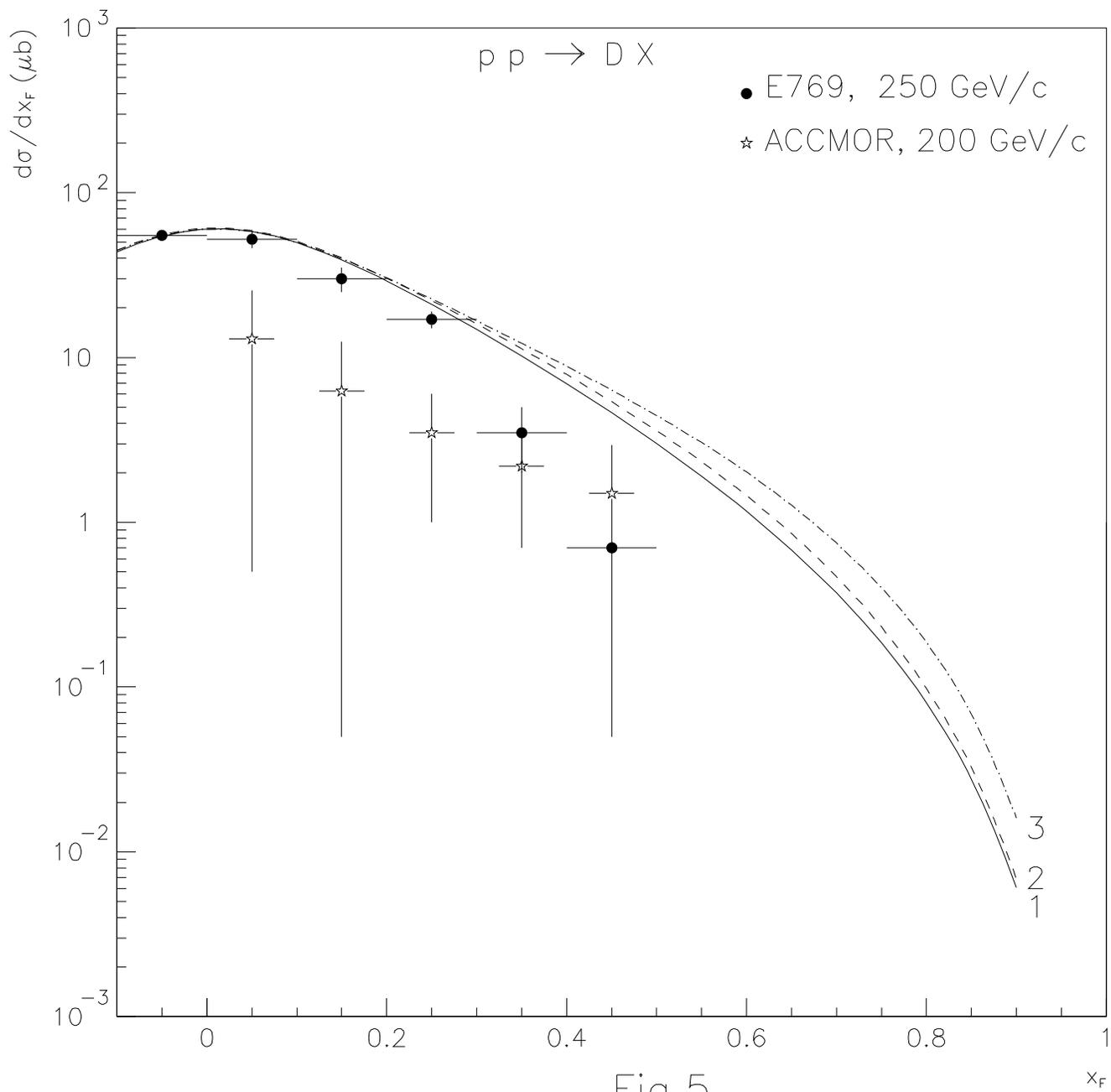
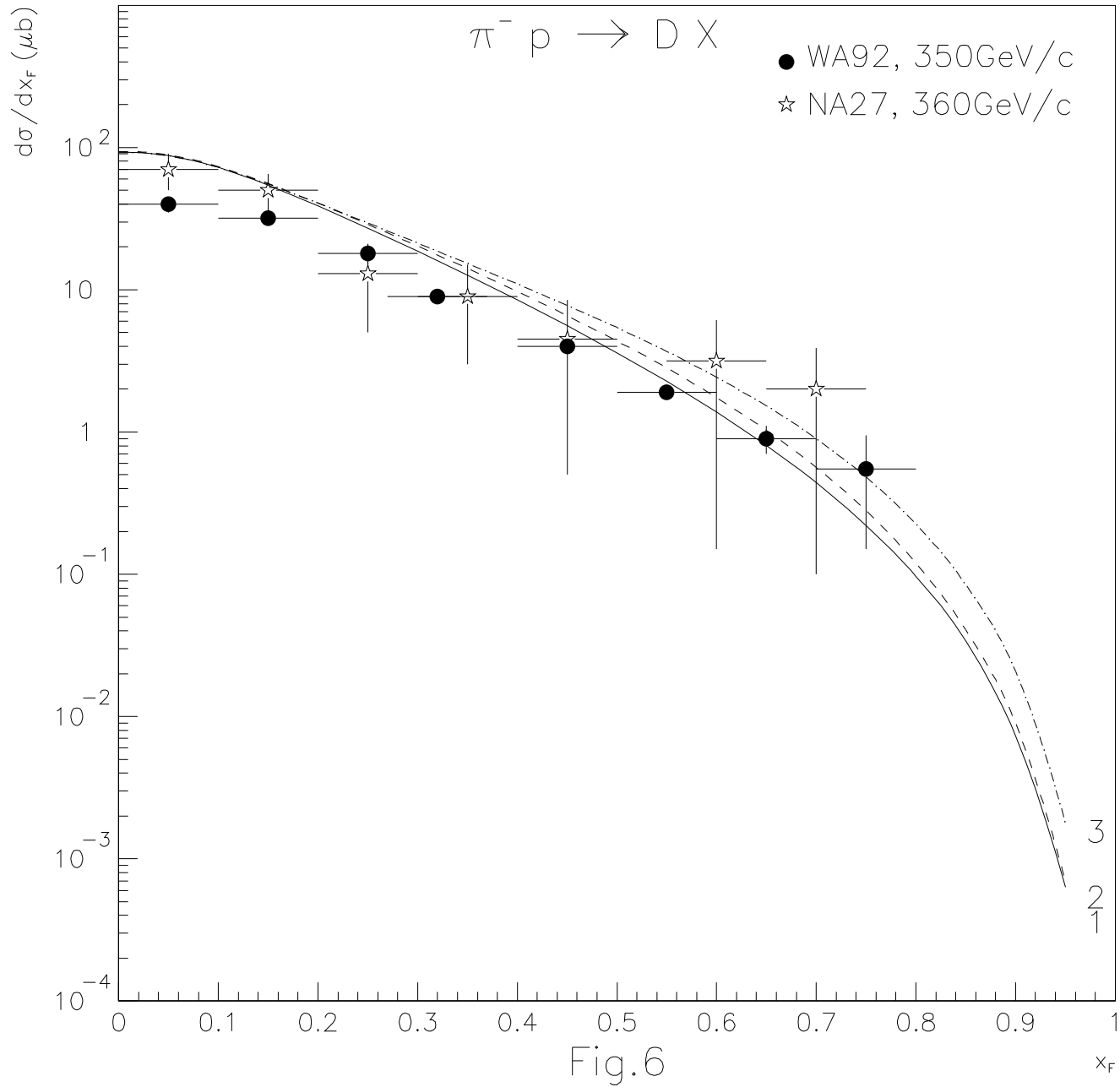
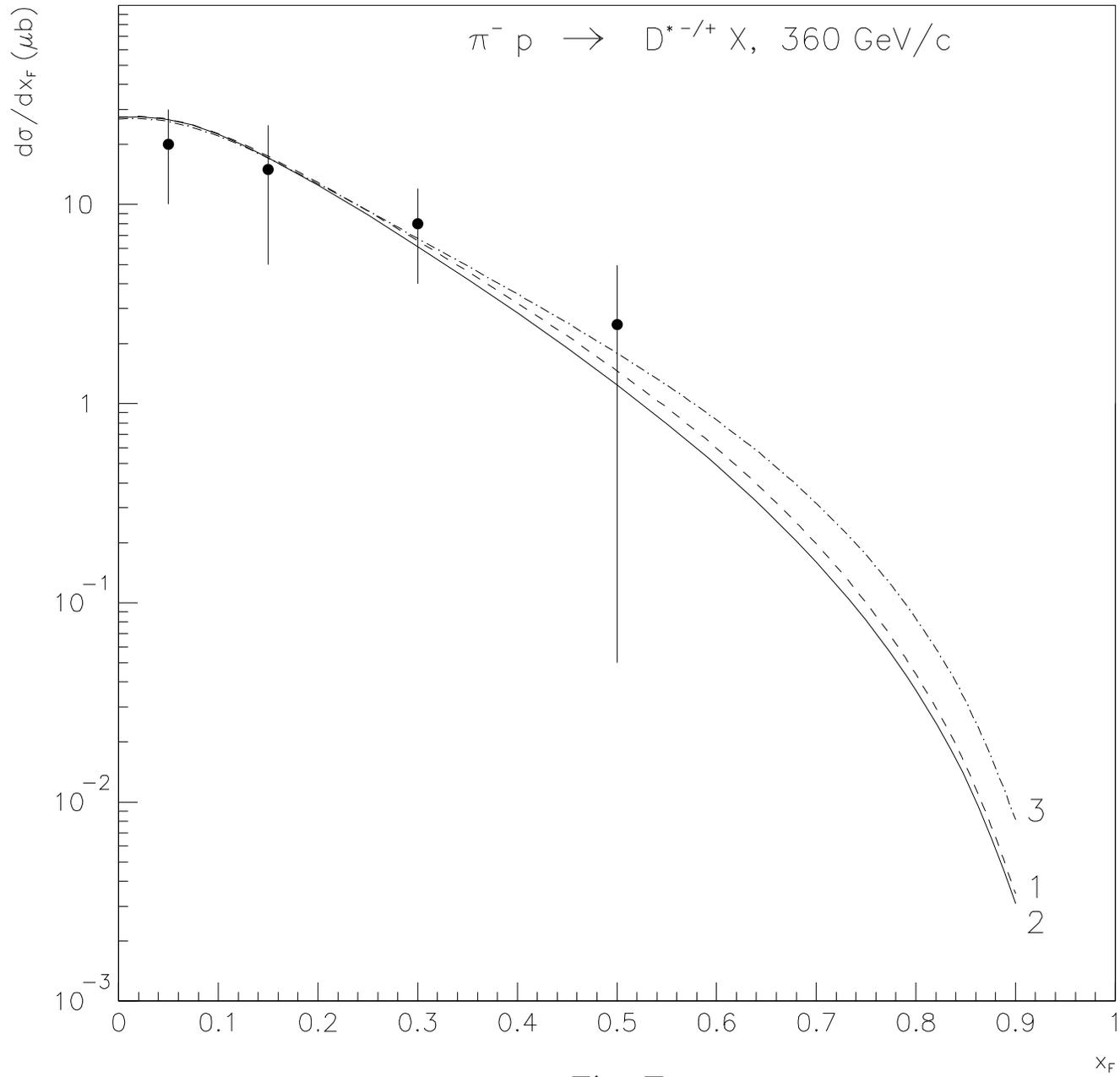
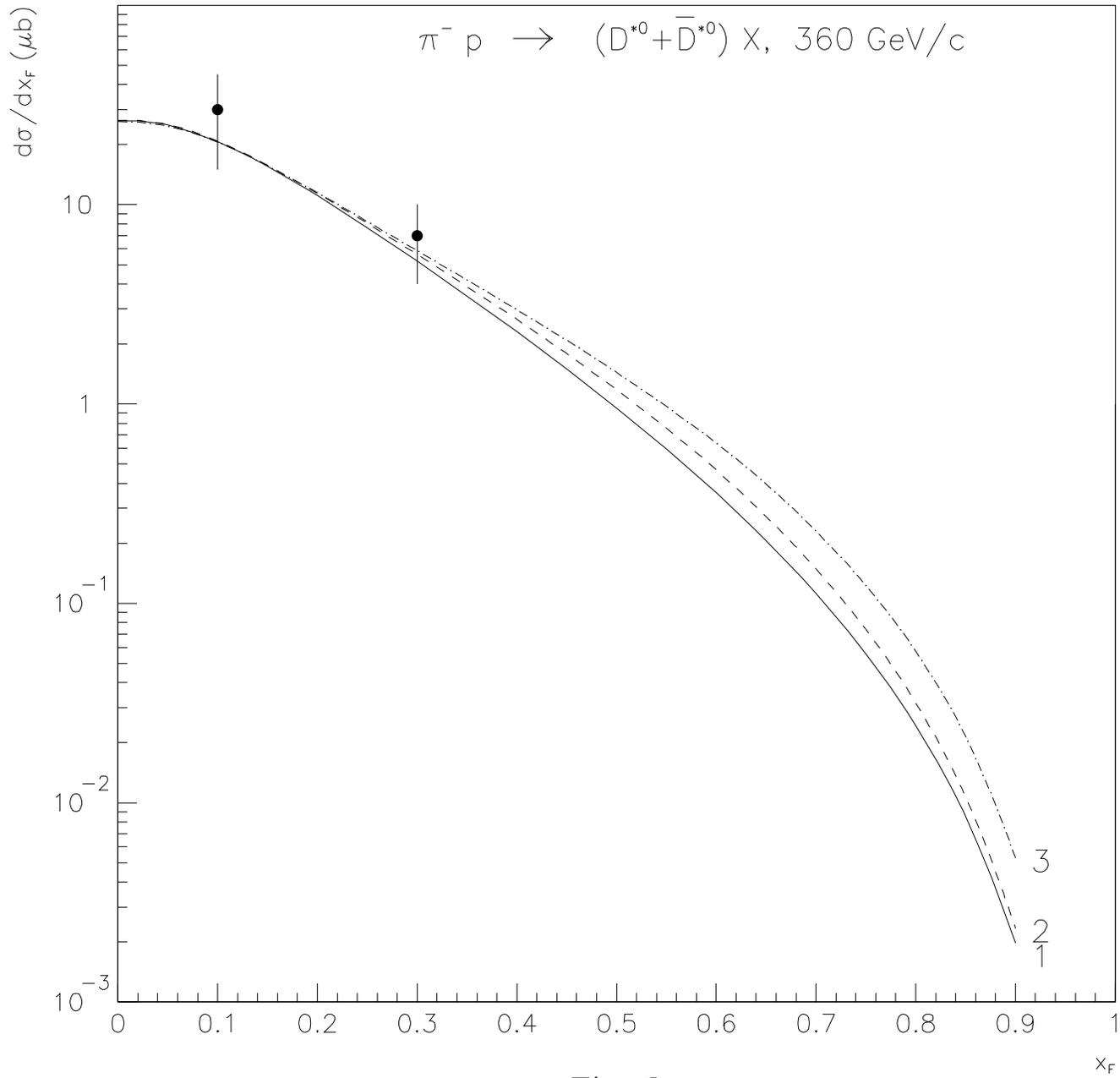
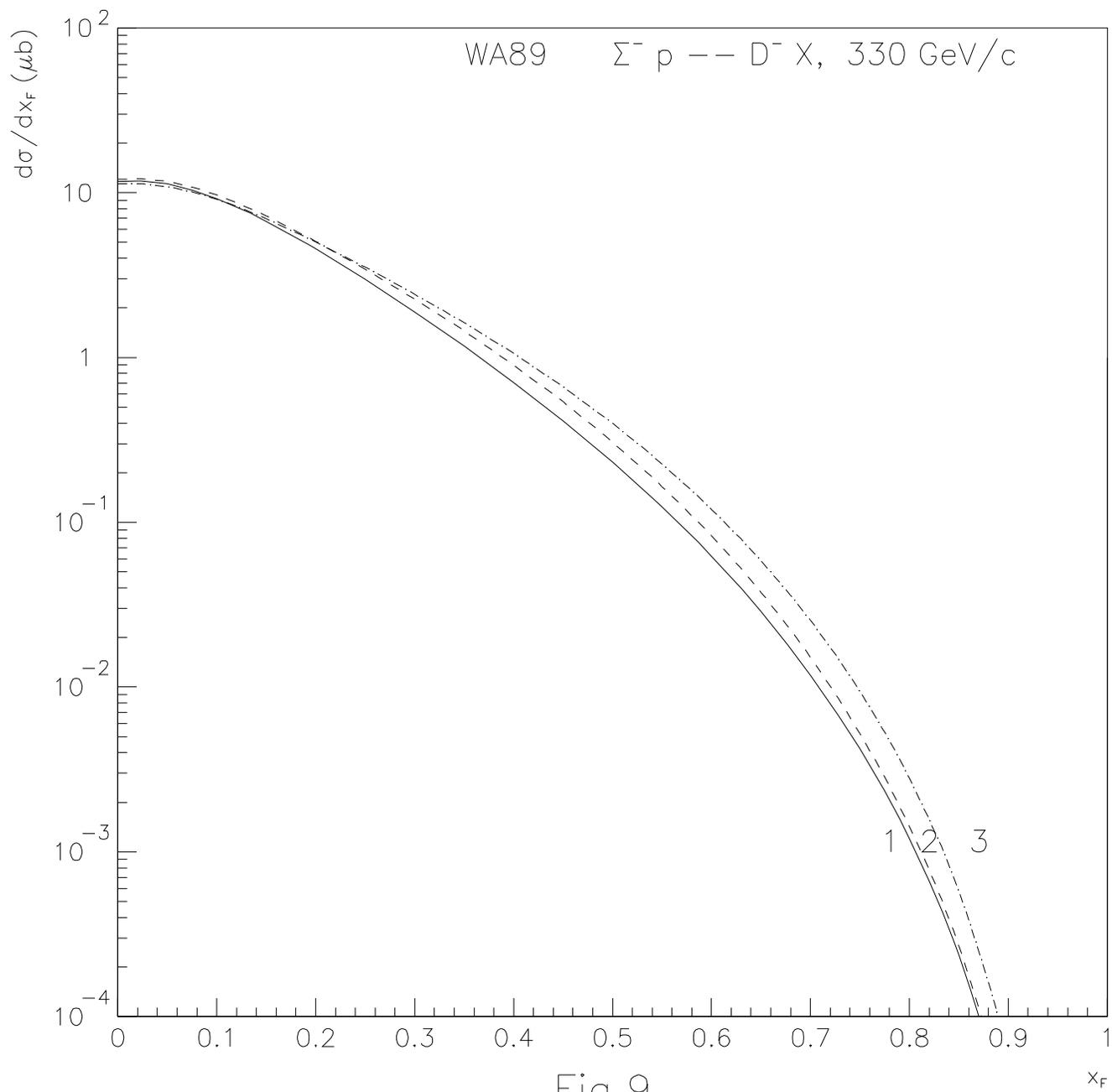
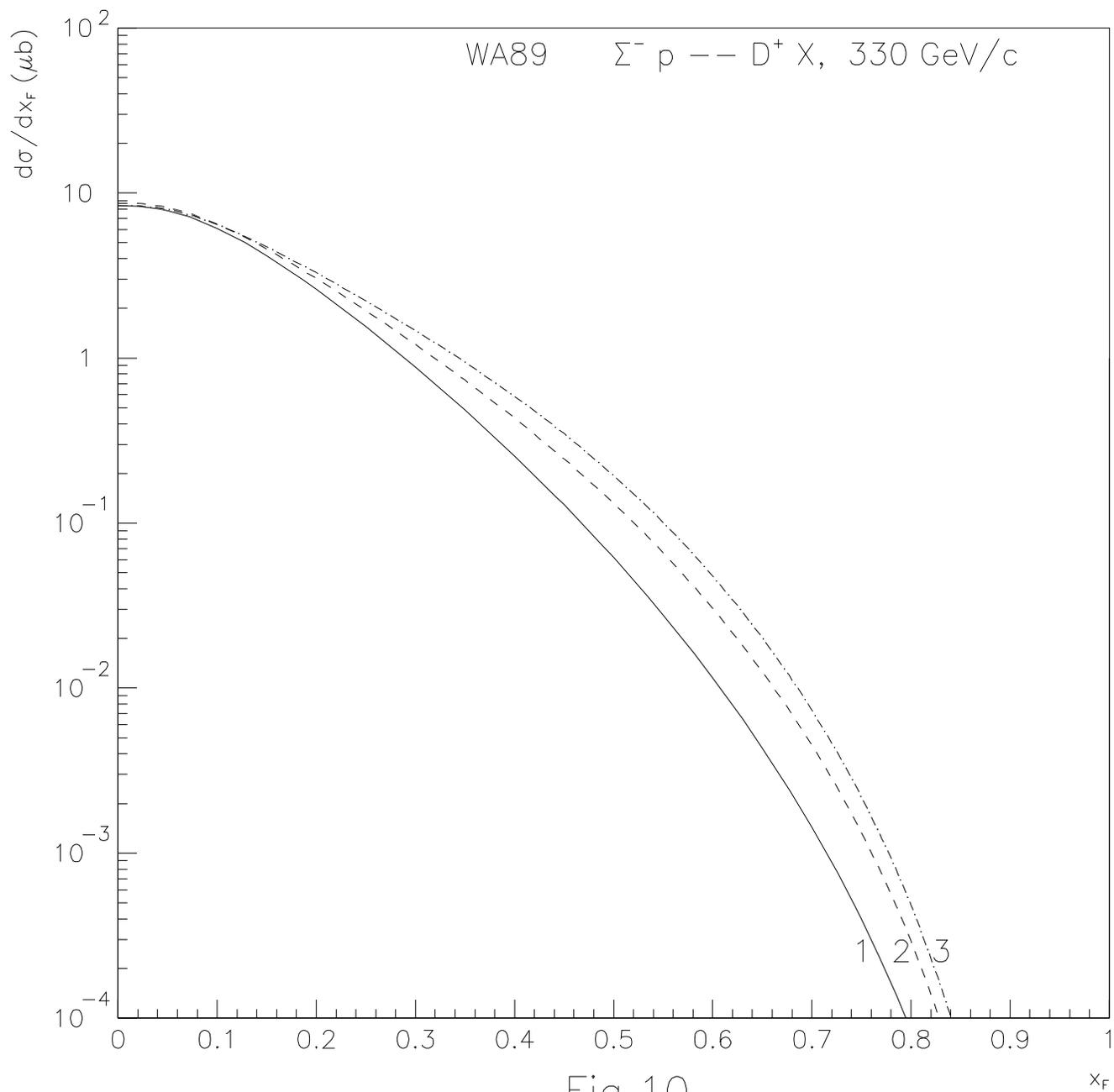
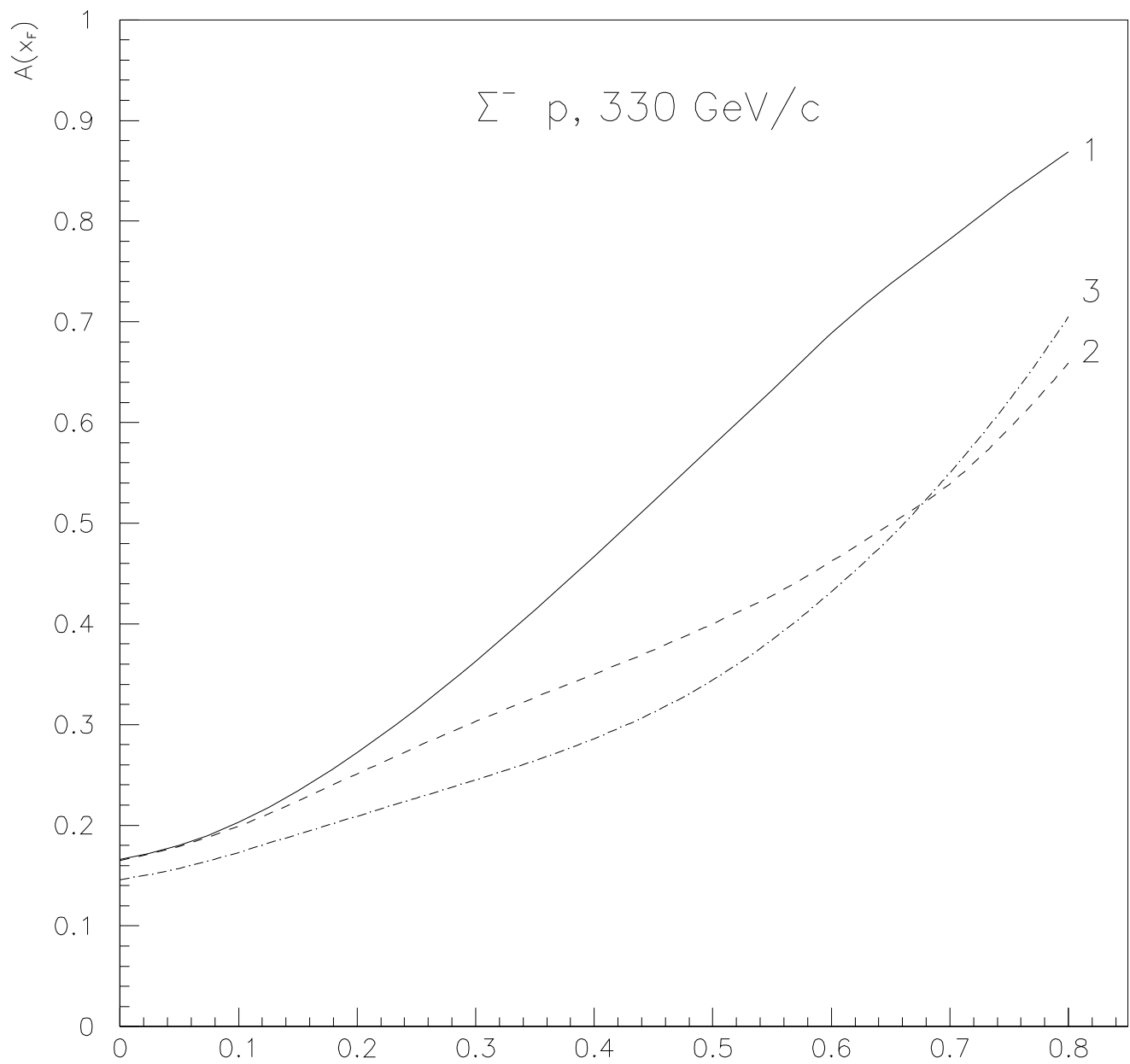
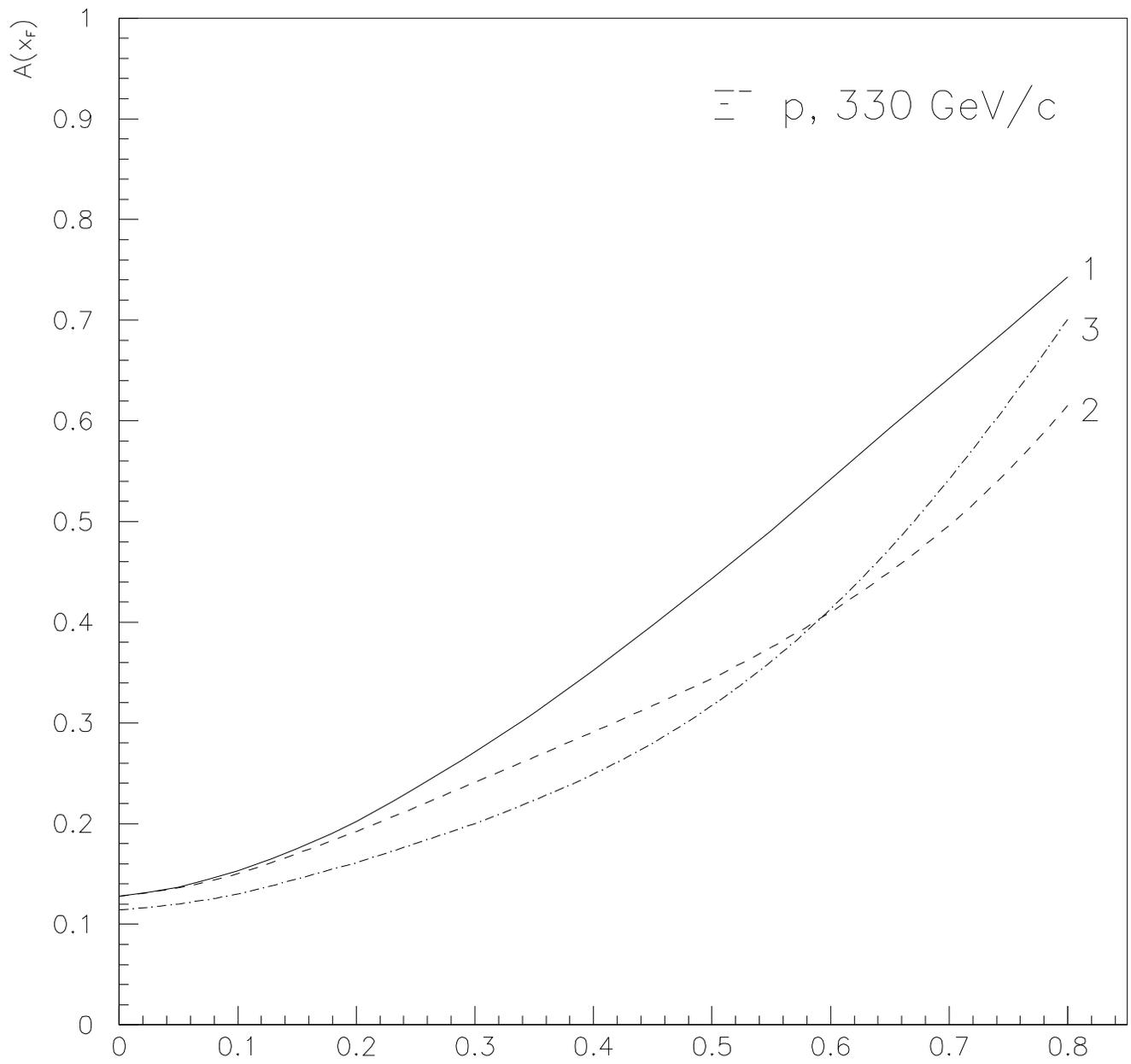
\end{document}